\documentclass{elsart}
\usepackage{graphics,natbib,graphicx,psfig,amssymb} 
\journal{New Astronomy}
\input psfig.sty
\begin{document}
\begin{frontmatter}
\title{Estimation of Galactic Model Parameters and Metalicity Distribution in Intermediate Latitudes with SDSS}
\author[istanbul]{E. Yaz\corauthref{cor}},
\corauth[cor]{corresponding author.}
\ead{esmayaz@yahoo.com}
\author[beykent]{S. Karaali}
\address[istanbul]{Istanbul University, Faculty of Sciences, Department of Astronomy and Space Sciences, 34119 University, Istanbul, Turkey}
\address[beykent]{Beykent University, Faculty of Science and Letters, Department of Mathematics and Computing, 
          Beykent 34398, Istanbul, Turkey}

\begin{abstract}
We estimated the galactic model parameters for a set of 20 intermediate-latitude fields 
with galactic longitudes $0^{\circ}\leq l \leq100^{\circ}$ and $160^{\circ}\leq l \leq240^{\circ}$, 
included in the currently available Data Release 6 (DR6) of the Sloan Digital Sky Survey 
({\em SDSS}), to explore their possible variation with galactic longitude. The local space 
densities of the thick disc and halo are almost the same for all fields, $\langle (n_{2}/n_{1})\rangle=6.52\%$ and 
$\langle (n_{3}/n_{1})\rangle=0.35\%$, respectively, a result different than the one cited for 
high-latitude fields. The thin disc's scaleheight is 325 pc in the galactic centre changes to 
369 pc in the third quadrant, which confirms the existence of disc flare, whereas the thick disc 
scaleheight is as large as 952 pc at galactic longitude $l=20^{\circ}$ and $10\%$ lower at 
$l=160^{\circ}$, which confirms the existence of the disc long bar in the direction 
$l\simeq27^{\circ}$. Finally, the variation of the axis ratio of the halo with galactic longitude 
is almost flat, $\langle(c/a)\rangle=0.56$, except a slight minimum and a small maximum in the 
second and third quadrants, respectively, indicating an effect of the long bar which seems 
plausible for a shallow halo. We estimated the metallicities of unevolved G--type stars and discussed 
the metallicity gradient for different vertical distances. The metallicity gradient is $d[M/H]/dz\sim-0.30$ 
dex kpc$^{-1}$ for short distances, confirming the formation of this region of the Galaxy by 
dissipational collapse. However, its change is steeper in the transition regions of different galactic 
components. The metallicity gradient is almost zero for inner halo ($5<z<10$ kpc), indicating a 
formation of merger or accretion of numerous fragments such as dwarf galaxies.

98.35.Ln Stellar content and populations; morphology and overall structure, 98.35.Gi Galactic halo

\end{abstract}

\end{frontmatter}

\section{Introduction}

Our knowledge of the structure of the Galaxy, as deduced from star count data with colour information, 
is more precise nowadays with the advent of new surveys. Researchers have used different 
methods to determine the galactic model parameters (cf. \citealt{Bilir06a}, Table 1). One can see that 
there is an improvement for numerical values of the model parameters. The local space density and 
the scaleheight of the thick disc can be given as an example. The scaleheight of the thick disc has steadily 
changed over the years from 1.45 to 0.65 kpc (\citealt{GR83, C01}) and higher local densities 
(2 - 10\%). \citet{J08} studied {\em SDSS} data and obtained 12\% relative local space 
density and 900 pc scaleheight for the thick disc. The authors assumed that 35\% of the sample stars were binaries.

In many studies the range of values for galactic model parameters is large. For example, \citet{C01}, 
\citet{Siegel02} give the local space density of the thick disc 6.5 - 13\% and 6 - 10\%, respectively. 
However, one expects the most evolved numerical values from these recent works. That is, either the 
range for this parameter should be small or a single value with a small error should be given for it. 
It seems that researchers have not been able to choose the most appropriate procedures for this topic. 
Large range or different numerical values for a specific galactic model parameter as estimated by 
different researchers may be due to several reasons: (a) The galactic model parameters are absolute 
magnitude dependent (\citealt{KBH04, Bilir06a}). Hence, any procedure which excludes this argument galactic 
model parameters spread in a large range. (b) The galactic model parameters are galactic latitude/longitude 
dependent. The two works of \citet{Buser98, Buser99} confirm this suggestion. Although these authors give a 
mean value for each parameter, the values of a given parameter are not equal for different fields. Several 
recent studies (\citealt{Bilir06b, Bilir06c, Cabrera-Lavers07, Ak07a, Bilir08a}) also support this argument. 
(c) Galactic model parameters change with limiting distance of completeness. That is, a specific model parameter 
is not the same for each set of galactic model parameters estimated for different volumes \citep{Karaali07}.

The difference between disc galactic model parameters estimated for fields with different galactic latitudes 
and longitudes are due to the influence of the disc flaring and warping. The disc of the Galaxy 
is far from being radially smooth and uniform. On the contrary, its overall shape presents strong asymmetries. 
While the warp bends the galactic plane upwards in the first and second galactic longitude quadrants 
($0 < l \leq180^{\circ}$) and downwards in the third and fourth quadrants ($180 < l \leq360^{\circ}$), the 
flare changes the scaleheight as a function of radial distance.

This warp is present in all galactic components: dust, gas and stars (\citealt{Lopez02, Momany06}). The stellar 
and gaseous flarings for the Milky Way are also consistent with each other \citep{Momany06}, showing that the scaleheight 
of the thin disc ($H_{1}$) increases with the galactocentric radius for $R>5$ kpc (\citealt{Kent91, DS01, 
NJ02, Lopez02, Momany06}). The behaviour of this flare in the central discs of spiral galaxies has not been 
studied so well due to inherent difficulties in separating the several contributions to the observed counts 
or flux. \citet{Lopez04} found that there is a deficit of stars compared to the predictions of a 
pure exponential law in the inner 4 kpc of the Milky Way, which could be explained as a flare, which 
displaces the stars to greater heights above the plane as the distance to the galactic centre decreases.

In this scenario, the mean disc ($z$=0) can be displaced as much as 2 kpc between the location of the 
maximum and the minimum amplitudes of the warp (\citealt{DS01, Lopez02, Momany06}). On the other hand, 
the scaleheight of the stars can show differences up to 50\% of the value for $H_{1}$ in the range 
$5<R<10$ kpc (\citealt{Al00, Lopez02, Momany06}) to fit a global galactic disc model that accounts 
for all these inhomogeneities is, at the very least, tricky. Because of this, the results in the galactic 
model parameters might depend on the sample of galactic coordinates used, as the combined effect of the warp 
and flare will be different in every direction in the Galaxy and hence along different lines of sight.

There is an additional reason for the difference between the numerical values of a given galactic model 
parameter estimated in different directions of the Galaxy, mainly at greater galactocentric distances. 
These are the observed overdense regions with respect to an axisymmetric halo, for which two competing 
scenarios have been proposed: the first one is concerned  with the triaxiality of the halo 
(\citealt{Newberg06, Xu06, J08}), whereas the second one is related to the remnants of some historical 
merger events \citep{Wyse05}. 

The size of the field used for galactic model estimation changes from author to author. There are star 
fields with sizes ranging from a few square degrees (cf. \citealt{Buser98, Buser99}) to fields with sizes 
larger than one thousand square degrees. For example the size of the field investigated by \citet{J08} 
is 6500 deg$^{2}$. Small sizes may involve some gaps causing unreliable galactic model parameters. On the 
other hand, very large fields may be contaminated by remnants of mergers. Also, as the galactic model 
parameters are galactic latitude and galactic longitude dependent, a specific model parameter estimated 
from investigation of a large field can be assumed as the mean of corresponding model parameters 
estimated for many neighbour small fields. 

Most of the {\em SDSS} data used in the works claimed above for galactic model estimation are reduced for high 
latitude fields. In this paper, we dealt with  the data of  intermediate latitude fields. We fixed the 
size of the fields with 10 deg$^{2}$, thus excluding the effect of the field size on the galactic models. 
Also, we avoided the contamination of mergers' remnants. So, we used recent {\em SDSS} Data Release 6 (DR6) 
of 20 intermediate latitude fields of equal size with galactic longitudes $0^\circ \leq l \leq 100^\circ$ 
and  $160^\circ\leq l \leq 240^\circ$ and estimated galactic model parameters for thin and thick discs and 
halo and compared them with the ones estimated for high latitude fields. Such a procedure gave us the chance 
to compare the trends of the galactic models estimated at different galactic latitudes and longitudes, as 
well as their interpretations. 

In Section 2, data and reductions are presented. The galactic model parameters and their dependence on 
galactic longitude and metallicity are given in Section 3. Section 4 provides a summary and discussion. 

\section{Data and Reductions}
The {\em SDSS\/} is a large, international collaboration project set up to survey 10 000 square-degrees of 
the sky in five optical passbands and to obtain spectra of one million galaxies, 100 000 quasars, and tens of 
thousands of galactic stars. The data are being taken with a dedicated 2.5-m telescope located at Apache 
Point Observatory (APO), New Mexico. The telescope has two instruments: a CCD camera with 30 CCD chips, each 
with a resolution of 2048$\times$2048 pixels, totaling approximately 120 million pixels, in the focal plane 
and two 320 fiber double spectrographs. The imaging data are tied to a network of brighter astrometric 
standards through a set of 22 smaller CCDs in the focal plane of the imaging camera. A 0.5-m telescope 
at APO has been used to tie the imaging data to brighter photometric standards. 

The {\em SDSS\/} obtains images almost simultaneously in five broad bands ($u$, $g$, $r$, $i$ and $z$)\footnote
{Magnitudes in this paper are quoted in the $ugriz$ system to differentiate them from the former $u^{'}g^{'}r^{'}i^{'}z^{'}$ system.}
centred at 3551, 4686, 6166, 7480 and 8932 $\AA$ \citep{Fukugita96} up to an apparent magnitude of 24.4, 25.3, 
25.1, 24.4 and 22.9, respectively, with a signal to noise ratio of 5 \citep{Adelman08}. The data becomes saturated at 
about 14 mag in $g$, $r$ and $i$ and about 12 mag in $u$ and $z$. The imaging data are automatically processed 
through a series of software pipelines which find and measure objects and provide photometric and astrometric 
calibrations to produce a catalogue of objects with calibrated magnitudes, positions and structure information. 
The photometric pipeline \citep{Lupton01} detects the objects, matches the data from the five filters, and 
measures instrumental fluxes, positions, and shape parameters (which allows the classification of objects as ``point source'', 
``compatible with the point spread function'' or ``extended''). The {\em SDSS\/} DR6 data can be obtained 
by using the SQL interface and includes the complete imaging of the Northern Galactic Cap. The catalogue contains images 
and parameters of 287 million objects over 9 583 deg$^2$, and 1.27 million spectra of sources (star, galaxy, quasar, etc.) 
over 7 425 deg$^2$ including scans over a large range of galactic latitudes and longitudes. The photometric calibration 
has been improved with uncertainties of 2\% in u and 1\% in $g$, $r$, $i$ significantly better than previous data releases 
\citep{Adelman08}.

\subsection{The Sample}
The data used in this work were taken from the {\em SDSS\/} DR6 WEB server\footnote {http://www.sdss.org/dr6/
access/index.html} for 22 intermediate-latitude fields ($44^\circ.3\leq b \leq45^\circ.7$) covering different 
galactic longitude intervals ($0^\circ< l \leq250^\circ$). Intermediate star fields with longitudes larger than 
$l=250^\circ$ were not available due to the {\em SDSS} observing strategy. {\em SDSS\/} magnitudes $u$, $g$, $r$, 
$i$ and $z$ were used for a total of 2$\times$10$^{6}$ stars in 22 fields. Although the fields are equal 
in size (10 deg$^{2}$), their surface densities (number of stars per square-degree) are not the same, following 
a specific trend with galactic longitude (Fig. 1). This is the first clue of galactic model parameters depending 
on galactic longitude. Owing to the {\em SDSS\/} observing strategy, data of stars brighter than $g_{0}=14$ mag, 
dereddened apparent $g$ magnitude, are saturated, and star counts are not complete for magnitudes fainter than 
$g_{0}= 22.2$ mag. Hence, our work was restricted to the magnitude range $15 < g_{0}\leq 22$ for the evaluation 
of the galactic model parameters. 

\begin{figure}
\begin{center}
\includegraphics[scale=0.45, angle=0]{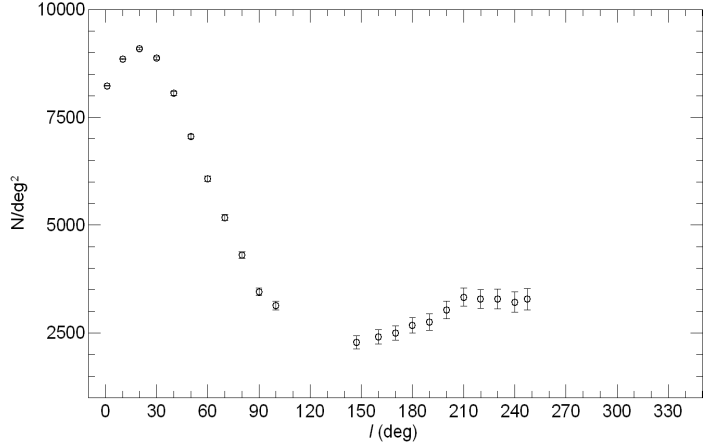}
\caption[]{Star counts at $\langle b \rangle=+45^\circ$ for 22 fields available in DR6.} 
\label{Fig01}
\end{center}
\end{figure}

\subsubsection{De-reddening of the Magnitudes}

The $E(B-V)$ colour excess was evaluated individually for each subsample source by using the maps of \citet{Schlegel98} 
through {\em SDSS} query server, $E(B-V)$ was reduced to total absorption $A_{V}$:

\begin{eqnarray}
A_{V}=3.1E(B-V).
\end{eqnarray}

In order to determine total absorptions for the {\em SDSS\/} bands ($A_{m}$), $A_{m}/A_{V}$ data given by \cite{Fan99}, 
i.e. 1.593, 1.199, 0.858, 0.639 and 0.459 for $m$ = $u$, $g$, $r$, $i$ and $z$, respectively, were used. Thus, the 
de-reddened magnitudes, with subscript 0, are  
 
\begin{eqnarray}
u_{0}=u-A_{u},\\
g_{0}=g-A_{g},\\
r_{0}=r-A_{r},\\
i_{0}=i-A_{i},\\
z_{0}=z-A_{z}.
\end{eqnarray}
The total absorptions $A_{m}$ are available in the {\em SDSS\/} query server.

All the colours and magnitudes mentioned hereafter will be de-reddened ones. Given that the location of 
the vast majority of our targets are at distances larger than 0.4 kpc, it seems appropriate to apply the 
full extinction from the maps \citep{Bilir08b}. Actually, when we combine the distance $r$=0.5 kpc ($z\sim$0.4 kpc 
distance from the galactic plane) with the scaleheight of the dust, \citep[$H$=125 pc]{Mar06}, we find that the total 
extinction is reduced to almost 6\% of the value galactic plane.

\subsubsection{The Star Sample}

According to \citet{C01}, the distribution of stars in the $g_{0}/(g-r)_{0}$ colour-magnitude diagram 
(CMD) can be classified as follows: Blue stars in the magnitude range $15<g_{0}<18$ are dominated by thick disc 
stars with a turn-off at $(g-r)_{0} \sim 0.33$, while galactic halo stars become significant for 
$g_{0}>18$, with a turn-off at $(g-r)_{0} \sim 0.2$; red stars ($(g-r)_{0} \geq 1.3$), are dominated 
by thin disc stars at all apparent magnitudes.    

However, the CMDs and the two-colour diagrams for all objects indicate that the stellar distributions are 
contaminated by extragalactic objects as claimed by \cite{C01}. The star/extragalactic object separation 
is based on the ``stellarity parameter'' as returned from the SE{\tiny XTRACTOR} routines \citep{Bertin96}. 
This parameter has a value between 0 (high extended) and 1 (point source). The separation works very well 
to classify a point source with a value greater than 0.8. Needless to say, this separation depends strongly 
on seeing and sky brightness. \cite{J08} applied an extra procedure in order to remove hot dwarfs, low-redshift 
quasars, and white/red dwarf unresolved binaries from their sample. This procedure consists of rejecting 
objects at distances larger than 0.3 mag from the stellar locus, i.e. the relation between $(g-r)_{0}$ and 
$(r-i)_{0}$ colours:

\begin{eqnarray}
(g-r)_{0}=1.39(1-\exp[-4.90(r-i)^{3}-2.45(r-i)^{2}-1.68(r-i)\\
-0.05]).\nonumber
\end{eqnarray}

\begin{figure}
\begin{center}
\includegraphics[scale=0.45, angle=0]{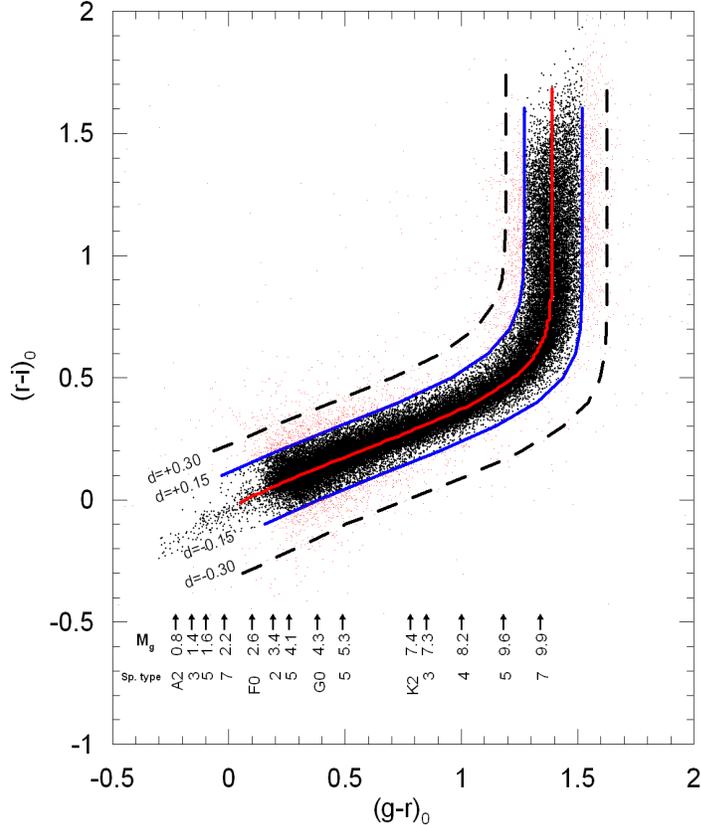}
\caption[]{Distribution of 60 960 point sources in the $(r-i)_{0}/(g-r)_{0}$ two-colour diagram. Thick solid line denotes \cite{J08}'s 
stellar locus. Thin solid and dashed lines represent the distance $d=\pm 0.15$ and $d=\pm 0.30$ mag from the stellar locus (Eq. 7 in the test), 
respectively. Also, absolute magnitudes and spectral types are given in the figure, confronted to $(g-r)_{0}$ colours.} 
\label{Fig02}
\end{center}
\end{figure}

The procedure of \cite{J08} works well for high latitude fields (cf. \citealt{Bilir08a}). However, it needs 
to be refined a bit for lower galactic latitude fields. We used Hipparcos' logarithmic local space density of stars 
with $4<M_{g}\leq10$, i.e. $D^{*}=7.49$ (see section 2.1.5), as a constraint and revealed that the distance $d=\pm 0.15$ mag is more appropriate 
than the distance $d=\pm 0.30$ mag for the fields with galactic latitude $\langle b \rangle=45^{\circ}$. Fig. 2 gives the 
$(g-r)_{0}/(r-i)_{0}$ two-colour diagram, the loci and the borders of the star sample in the star field with 
longitude $l=60^{\circ}$. The $g_{0}/(g-r)_{0}$ CMD for the same field in Fig. 3 confirms the arguments of \cite{C01}. 

\begin{figure}
\begin{center}
\includegraphics[scale=0.45, angle=0]{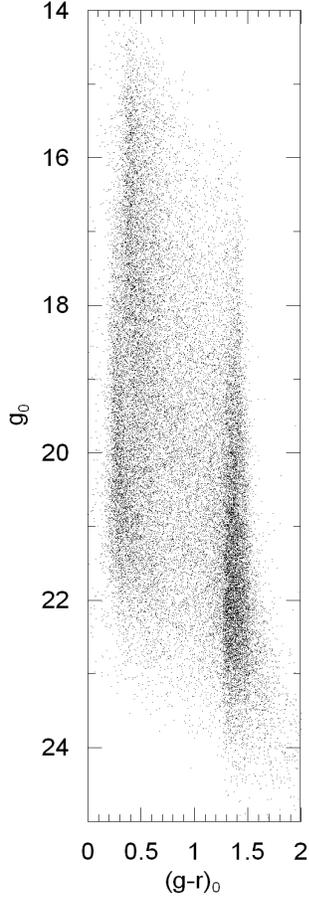}
\caption[]{$g_{0}/(g-r)_{0}$ colour-magnitude diagram for stars in the field with galactic longitude $l=60^\circ$.}
\label{Fig03}
\end{center}
\end{figure}

The histogram of the apparent magnitudes of the star sample in Fig. 4 shows that the sample is complete for stars with  
apparent magnitudes $15<g_{0}\leq22$. Hence, the estimation of the galactic model parameters is based on these data.

\begin{figure}
\begin{center}
\includegraphics[scale=0.45, angle=0]{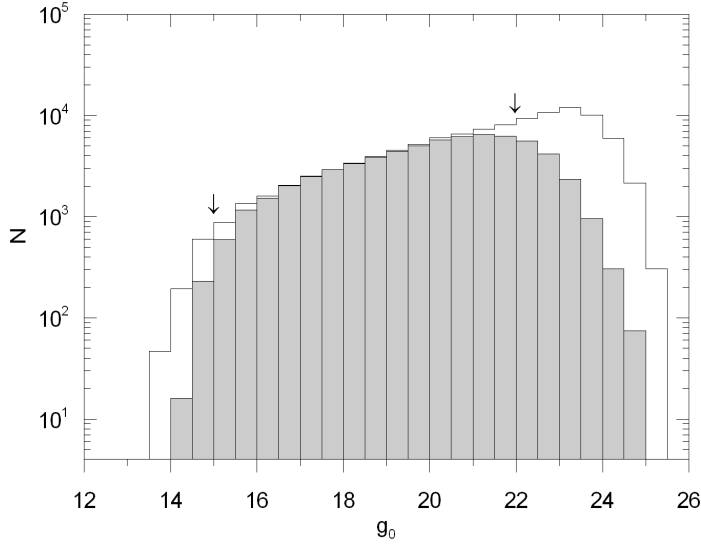}
\caption[]{Apparent magnitude histogram for point sources (white area) and star sample (shaded area) for the field 
centred at $l=60^\circ$.}
\label{Fig04}
\end{center}
\end{figure}

\subsubsection{Dwarf--Gaint Separation}

We used the selection criteria of \cite{Helmi03} to separate the metal-poor giants. These authors define the 
location of the metal-poor giants by the following criteria: $r_{0}<19$ mag, $1.1\leq(u-g)_{0}\leq 2.0$, $0.3\leq(g-r)_{0}\leq0.8$, 
$0.1<P_{1}<0.6$, $\mid s\mid> m_{s}+0.05$, where $P_{1}=0.910(u-g)_{0}+ 0.415(g-r)_{0}-1.28$, $s=0.249u_{0} + 0.794g_{0} - 0.555r_{0} + 0.24$ 
and $m_{s}$=0.002. When we apply these criteria to the star sample we obtain $\sim$ 5 069 metal--poor giants which corresponds to 0.6\% of 
the whole sample. The projection of the location of the dwarfs and giants on the $(g-r)_{0}/(r-i)_{0}$ two-colour diagram 
for the field with longitude $l=60^{\circ}$ is given in Fig. 5.   

\begin{figure}
\begin{center}
\includegraphics[scale=0.40, angle=0]{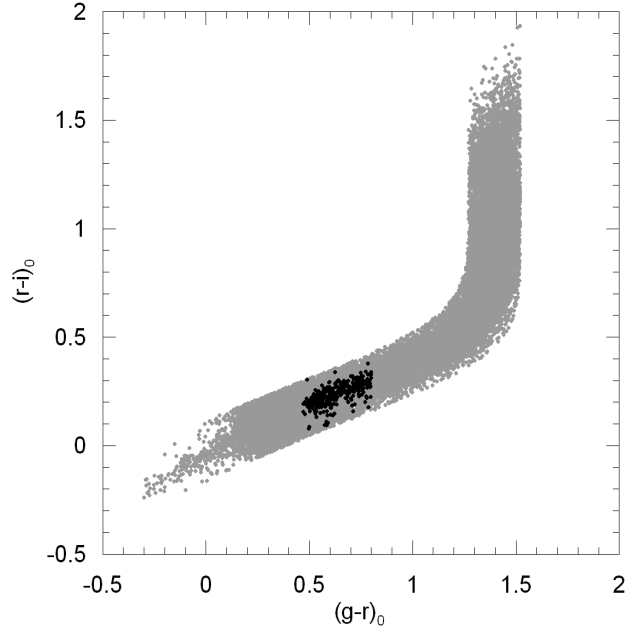}
\caption[]{Metal-poor giant stars (black dots) in the $(r-i)_{0}/(g-r)_{0}$ two-colour diagram for stars in the field centred at $l=60^\circ$.}
\label{Fig05}
\end{center}
\end{figure}             

As we omitted the bright stars ($g_{0}<15$ mag) due to saturation, we do not expect a sufficient number of metal-rich stars, 
affecting our statistics. Thus, the total number of dwarfs in the sample is 854 960.

\subsubsection{Absolute Magnitudes and Distances}

Absolute magnitudes were determined using two different procedures. For absolute magnitudes $4<M_{g}\leq8$ we used 
the procedure of \citet{KBT05}, whereas for $8<M_{g}\leq10$ we adopted the procedure of \citet{Bilir05}. 
In the procedure of KBT, the absolute magnitude offset from the Hyades main sequence, $\Delta M_{g}^{H}$, is given 
as a function of both $(g-r)_{0}$ colour and $\delta_{0.43}$ UV-excess, as follows:

\begin{eqnarray}
\Delta M_{g}^{H}=c_{3}\delta^{3}_{0.43}+c_{2}\delta^{2}_{0.43}+c_{1}\delta_{0.43}+c_{0},
\label{eq8}
\end{eqnarray}
where $\delta_{0.43}$ is the UV-excess of a star relative to a Hyades star of the colour-index $(g-r)_{0}=0.43$ which 
corresponds to $\delta_{0.6}$ and is determined using the colour transformation equations between {\em UBV} and {\em SDSS} photometry (KBT). 
The coefficients $c_{i}$ ($i$=0, 1, 2, 3) are functions of $(g-r)_{0}$ colour and are adopted from the work of KBT, 
where $\Delta M^{H}_{g}$ is defined as the difference in absolute magnitudes of a program star and a Hyades star of 
the same $(g-r)_{0}$ colour:

\begin{eqnarray}
\Delta M^{H}_{g} = M^{*}_{g}-M^{H}_{g}.
\label{eq8}
\end{eqnarray}

The absolute magnitude of a Hyades star can be evaluated from the Hyades sequence, normalized by KBT. This procedure 
is the one used in the works of \cite{Ak07a}, \cite{Karaali07} and \cite{Bilir08a}, and has two main advantages: 1) 
there is no need to separate the stars into different populations, and 2) the absolute magnitude of a star is determined from 
its UV-excess individually which provides more accurate absolute magnitudes compared with the procedure ``{\it in-situ}'', 
 where a specific CMD is used for all stars of the same population. When one uses the last two equations (Eqs. 8 and 9) 
and the following one (Eq. 10), which provides absolute magnitudes for the Hyades stars it gets the 
absolute magnitude $M_{g}^{*}$ of a star: 

\begin{eqnarray}
M_{g}^{H} = -2.0987(g-r)^{2}-0.0008(u-g)^{2}+0.0842(g-r)(u-g)\\
+7.7557(g-r)- 0.1556(u-g)+1.9714.\nonumber
\end{eqnarray}

The procedure of KBT was defined for the colour range $0.09<(g-r)_{0}\leq0.93$, which corresponds to absolute magnitudes 
$4<M_{g}\leq8$. Hence, for the absolute magnitudes interval $8<M_{g}\leq10$ we used the equation of BKT, which provides absolute 
magnitudes for late-type dwarfs:
 
\begin{eqnarray}
M_{g} = 5.791 (g-r)_{0} + 1.242 (r-i)_{0} + 1.412.
\end{eqnarray}

Stars with faint absolute magnitudes are very useful, since they provide space densities at short 
distances relative to the Sun, which combine the space densities for bright stars at large distances, and 
the local densities of {\em Hipparcos} \citep{Jahreiss97}. Thus, we have a sample of stars with absolute magnitudes 
$4<M_{g}\leq10$, which enables us to evaluate space density functions in the heliocentric distance interval $0.5<r\leq30$ 
kpc, which corresponds to a range of distances of $0.4<z\leq20$ kpc from the galactic plane. This interval is large enough 
to estimate a set of galactic model parameters and test their changes with galactic longitude. The absolute magnitudes 
in question and the corresponding spectral types (from early F type to early M type) for the locus points in the 
$(g-r)_{0}/(r-i)_{0}$ two--colour diagram are shown in Fig. 2 for the field centred at $l=60^\circ$ as an example. 
 The local space density in the absolute magnitude interval $4<M_{g}\leq10$ is flat and it attributes to a mean value of 
logarithmic space density $D^{*}=7.49$.

In a conical magnitude-limited volume, the distance up to which intrinsically bright stars are visible is larger than 
the distance up to which intrinsically faint stars are visible. This causes brighter stars to be statistically 
overrepresented and the derived absolute magnitudes to be too faint. This effect, known as Malmquist bias \citep{M20}, 
was formalized into the general formula:

\begin{equation}
M(g)=M_{0}-\sigma^{2}{d\log A(g) \over dg},
\end{equation}

where $M(g)$ is the assumed absolute magnitude, $M_{0}$ is the absolute magnitude calculated for any star using KBT 
calibration, $\sigma$ is the dispersion of the KBT or BKT calibration, and $A(g)$ is the differential counts evaluated 
at the apparent magnitude $g_{0}$ of any star. The dispersion in absolute magnitude calibration of KBT and BKT is 
around 0.25 mag, corresponding to an error of 10\% in photometric distance. We divided the sample into the absolute magnitude 
intervals (4,5], (5,6], (6,7], (7,8], (8,9] and (9,10], and we applied the Malmquist bias to stars in each interval 
separately. This approach provides uniform space densities which is essentially the Malmquist bias. Thus, 
the corrections applied to the absolute magnitudes are 0.005, 0.003, 0.007, 0.008, 0.012 and 0.012 for the absolute 
magnitude intervals cited above. The correction of the Malmquist bias was applied to {\em SDSS} photometric data 
used in this work.

Combination of the absolute magnitude $M_{g}$ and the apparent magnitude $g_{0}$ of a star gives its distance $r$ 
relative to the Sun, i.e.

\begin{equation}
[g-M_{g}]_{0}=5\log r-5.
\end{equation}

\cite{Gilmore95} quote an error of $\sim 0.2$ dex in the derivation of $[M/H]$ from the {\it UBV} photometry for F/G stars 
which leads to a random uncertainty of 20\% in the distance estimation. One expects larger distance errors for late 
spectral type stars. However, $(u-g)$ and $(g-r)$ colours are more accurate than the $(U-B)$ and $(B-V)$ 
colours which mitigate this excess error for K stars. The distance of the star to the galactic plane can be evaluated 
using its distance $r$ and galactic latitude $b$:

\begin{eqnarray}
z=r \sin b.					
\end{eqnarray}

\subsubsection{Density Functions}

Logarithmic space densities $D^{*}=\log D+10$ have been evaluated for the combination of three population components 
(thin and thick discs and halo), for each field where $D=N/\Delta V_{1,2}$; $\Delta V_{1,2}=(\pi/180)^{2}(\square/3)(r_{2}^{3}-r_{1}^{3})$; 
$\square$ denotes the size of the field (10 deg$^{2}$); $r_{1}$ and $r_{2}$ are the lower and upper limiting distances of the volume 
$\Delta V_{1,2}$; $N$ is the number of stars per unit absolute magnitude; $r^{*}=[(r^{3}_{1}+r^{3}_{2})/2]^{1/3}$ is the 
centroid distance of the volume $\Delta V_{1,2}$; and $z^{*}=r^{*}\sin b$, $b$ being the galactic latitude of the field 
centre. The limiting distances of completeness $r_{l}$ and $z_{l}$ can be calculated by substituting $g_{l}$, $r_{l}$ and  
$z_{l}$ for $g$, $r$ and $z$ in  Eqs. 13 and 14, where $g_{l}$ is the limiting apparent magnitude (15 and 22 mag 
for bright and faint stars, respectively). The logarithmic density function evaluated for the field centred at the galactic 
longitude $l=60^{\circ}$ is given in Fig. 6.

\subsubsection{Density Laws}

In this work we adopted the density laws of Basle group \citep{Buser98, Buser99}. Disc structures are usually parametrized in 
cylindrical coordinates using radial and vertical exponentials:

\begin{equation}
D_{i}(x,z)=n_{i}~\exp(-|z|/H_{i})~\exp(-(x-R_{0})/h_{i}),\\
\label{ec1}
\end{equation}

where $z=z_{\odot}+r\sin b$, $r$ is the distance to the object from the Sun, $b$ is the galactic latitude, $z_{\odot}$ is the vertical
distance of the Sun from the galactic plane \citep[24 pc]{J08}, $x$ is the projection of the galactocentric distance on the galactic 
plane, $R_{0}$ is the solar distance from the galactic centre (8 kpc; \citealt{R93}), $H_{i}$ and $h_{i}$ are the scaleheight and 
scalelength, respectively, and $n_{i}$ is the normalized density at the solar radius. The suffix $i$ is 1 for the thin disc, whereas 
2 for the thick disc. 
 
The density law for the spheroid component is parameterized in different forms. The most common is the \citet{deV48} spheroid 
used to describe the surface brightness profile of elliptical galaxies. This law has been deprojected into three dimensions by 
\citet{Young76} as 

\begin{eqnarray}
D_{s}(R)=n_{s}~(\exp[10.093(1-R/R_{0})^{1/4}]/(R/R_{0})^{7/8})\\
(1-[0.08669/(R/R_{0}]),\nonumber
\end{eqnarray}
where $R$ is the (uncorrected) galactocentric distance in spherical coordinates, and $n_{s}$ is the normalized local density. $R$ 
has to be corrected for the axial ratio $(c/a)$, 
\begin{eqnarray}
R = [x^{2}+(z/(c/a))^2]^{1/2},
\end{eqnarray}
where
\begin{eqnarray}
z = r \sin b,
\end{eqnarray}
\begin{eqnarray}
x = [R_{0}^{2}+r^{2}\cos^{2} b-2R_{0}r\cos b \cos l]^{1/2}, 
\end{eqnarray}
with $r$ being the distance along the line of sight and, $l$, $b$ being the galactic longitude and latitude respectively, for the field 
under investigation.

\section{Galactic Model Parameters and Metallicity Distribution}
\subsection{Estimation of the Galactic Model Parameters}

We estimated all the galactic model parameters simultaneously, by fitting the space density functions derived from the 
observations (combined for the three population components) to a corresponding combination of the adopted population-specific 
analytical density laws. The faintest stars in this work provide space densities at short distances from the 
galactic plane, $z\sim0.4$ kpc. Hence, it is possible to have reliable extrapolation between them and the space density of 
{\em Hipparcos} in the solar neighbourhood, $D^{*}=7.49$ in logarithmic form \citep{Jahreiss97}, corresponding to the mean of 
the local space densities for stars with $4<M_{g}\leq10$.

We used the classical $\chi^{2}$ statistic to estimate galactic model parameters, which is the most commonly method 
in recent studies (\citealt{Phleps00, Phleps05, C01, Siegel02, Du03, Du06, J08, Bilir08a}). The comparison of the 
logarithmic density functions derived from the observations and the analytical density laws is given in Fig. 6 for the 
field with galactic longitude $l=60^{\circ}$, as an example. $\chi_{min}^{2}$ shows a symmetrical distribution, as it can be 
seen from Fig. 7, which is given as an example. Hence, the errors of the galactic model parameters could be estimated 
by changing a given model parameter until an increase or decrease by 1 was achieved \citep{Phleps00}. Table 1 lists the 
galactic model parameters for 20 {\em SDSS} intermediate-latitude ($44^{\circ}.3\leq b\leq45^{\circ}.7$) fields, resulting from the 
fits of the analytical density profiles (fields with galactic longitudes $l=150^{\circ}$ and $l=250^{\circ}$ are omitted from 
the table due to their small sizes, $\sim$4 deg$^{2}$). The columns indicate: galactic longitude ($l$), scaleheight of thin 
($H_{1}$) and thick discs ($H_{2}$), local space densities of the thick disc ($n_{2}/n_{1}$) and the halo ($n_{3}/n_{1}$) 
relative to the local space density of the thin disc, scalelength of the thin ($h_{1}$) and thick discs ($h_{2}$), axial 
ratio of the halo ($c/a$), reduced chi-square minimum ($\widetilde{\chi}^{2}_{min}$), standard deviation ($s$) and the 
corresponding probability. The $\widetilde{\chi}^{2}_{min}$ values are low, whereas the probabilities are rather high 
confirming the reality of the galactic model parameters.

\begin{figure}
\begin{center}
\includegraphics[scale=0.45, angle=0]{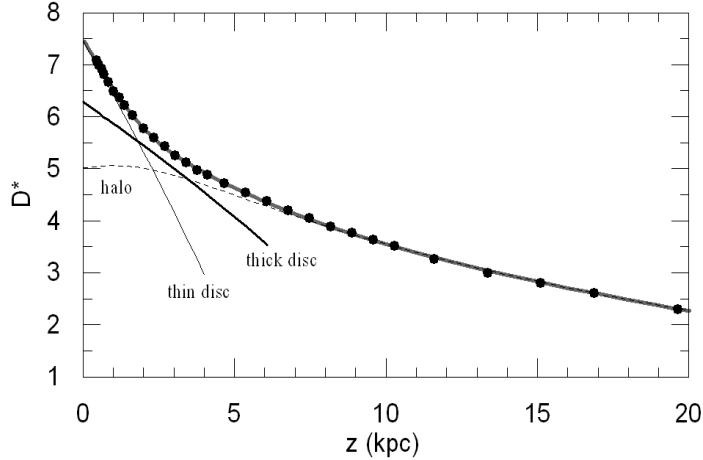}
\caption[]{Observed (dots) and evaluated (thick solid line) space density functions combined for stars of all three 
population components. Also, independent space density functions of thin disc (thin dashed line), thick disc (solid 
line) and halo (dotted line) for the star field with $l=60^\circ$ are shown.} 
\label{Fig06}
\end{center}
\end {figure}

\begin{table*}
\setlength{\tabcolsep}{4pt}
{\tiny
\center
\caption{Estimated galactic model parameters for 20 star fields.}
\begin{tabular}{cccccccccccc}
\hline
       $l$ &  $H_{1}$   &    $h_{1}$      &   $ H_{2}$    &       $h_{2}$    &  $(n_{2}/n_{1})$ &         (c/a)     &  $(n_{3}/n_{1})$ &$\widetilde{\chi}^{2}_{min}$ & s & Prob \\
         (deg)  & (pc)         & (kpc) & (pc)    &  (kpc)  & (\%) &       & (\%) & \\
\hline
         0 &  $325\pm6$ &  $1.00\pm0.39$  &   $946\pm26$  &   $4.48\pm0.39$  &   $6.56\pm0.63$  &   $0.545\pm0.01$  &  $0.33\pm0.02$  &    7.95 &   0.022 &   0.997 \\
        10 & $332\pm6$  &  $1.00\pm0.22$  &   $952\pm45$  &   $4.21\pm0.32$  &   $6.55\pm0.56$  &   $0.560\pm0.01$  &  $0.33\pm0.01$  &    4.36 &   0.017 &   0.999 \\
        20 & $343\pm7$  &  $1.00\pm0.43$  &   $946\pm25$  &   $5.49\pm0.74$  &   $6.50\pm0.96$  &   $0.565\pm0.01$  &  $0.33\pm0.01$  &    5.83 &   0.019 &   0.999 \\
        30 & $348\pm5$  &  $1.00\pm0.44$  &   $951\pm20$  &   $4.00\pm0.44$  &   $6.46\pm0.39$  &   $0.570\pm0.01$  &  $0.33\pm0.01$  &    7.92 &   0.023 &   0.997 \\
        40 & $352\pm6$  &  $1.00\pm0.53$  &   $941\pm24$  &   $3.67\pm0.49$  &   $6.49\pm0.44$  &   $0.569\pm0.01$  &  $0.33\pm0.01$  &    7.23 &   0.022 &   0.998 \\
        50 & $354\pm9$  &  $1.00\pm0.69$  &   $934\pm22$  &   $3.35\pm0.52$  &   $6.52\pm0.51$  &   $0.569\pm0.01$  &  $0.34\pm0.01$  &    8.26 &   0.023 &   0.996 \\
        60 & $352\pm6$  &  $1.05\pm0.12$  &   $931\pm22$  &   $3.52\pm0.94$  &   $6.52\pm0.49$  &   $0.564\pm0.01$  &  $0.39\pm0.01$  &    7.94 &   0.023 &   0.997 \\
        70 & $353\pm7$  &  $1.16\pm0.31$  &   $921\pm18$  &   $3.11\pm1.32$  &   $6.56\pm0.50$  &   $0.551\pm0.01$  &  $0.36\pm0.01$  &    9.31 &   0.025 &   0.991 \\
        80 & $350\pm7$  &  $1.90\pm0.77$  &   $902\pm18$  &   $2.90\pm1.53$  &   $6.52\pm0.43$  &   $0.537\pm0.01$  &  $0.36\pm0.01$  &    7.23 &   0.021 &   0.998 \\
        90 & $331\pm10$ &  $1.01\pm0.76$  &   $882\pm24$  &   $2.65\pm0.78$  &   $6.40\pm0.45$  &   $0.520\pm0.01$  &  $0.36\pm0.01$  &    7.27 &   0.022 &   0.998 \\
       100 & $330\pm9$  &  $1.36\pm0.47$  &   $883\pm17$  &   $3.90\pm0.91$  &   $6.47\pm0.38$  &   $0.510\pm0.01$  &  $0.36\pm0.01$  &    8.01 &   0.022 &   0.997 \\
       160 & $357\pm8$  &  $1.38\pm0.12$  &   $860\pm17$  &   $4.15\pm0.35$  &   $6.46\pm0.39$  &   $0.489\pm0.01$  &  $0.33\pm0.01$  &    8.10 &   0.023 &   0.996 \\
       170 & $355\pm9$  &  $1.48\pm0.14$  &   $863\pm17$  &   $4.15\pm0.39$  &   $6.55\pm0.39$  &   $0.510\pm0.01$  &  $0.34\pm0.01$  &    8.40 &   0.023 &   0.995 \\
       180 & $362\pm8$  &  $1.66\pm0.16$  &   $875\pm32$  &   $4.22\pm0.56$  &   $6.62\pm0.54$  &   $0.537\pm0.01$  &  $0.34\pm0.02$  &    7.00 &   0.020 &   0.999 \\
       190 & $363\pm8$  &  $1.65\pm0.25$  &   $877\pm17$  &   $4.50\pm0.44$  &   $6.62\pm0.43$  &   $0.551\pm0.01$  &  $0.35\pm0.02$  &    8.29 &   0.023 &   0.996 \\
       200 & $369\pm9$  &  $1.68\pm0.17$  &   $904\pm23$  &   $4.45\pm0.48$  &   $6.56\pm0.40$  &   $0.610\pm0.01$  &  $0.36\pm0.02$  &    6.97 &   0.021 &   0.999 \\
       210 & $368\pm9$  &  $1.68\pm0.18$  &   $905\pm20$  &   $4.47\pm0.50$  &   $6.59\pm0.40$  &   $0.654\pm0.02$  &  $0.39\pm0.02$  &    8.75 &   0.020 &   0.994 \\
       220 & $365\pm8$  &  $1.53\pm0.17$  &   $895\pm22$  &   $3.80\pm0.41$  &   $6.55\pm0.41$  &   $0.651\pm0.01$  &  $0.37\pm0.01$  &    7.74 &   0.022 &   0.997 \\
       230 & $360\pm9$  &  $1.23\pm0.13$  &   $899\pm20$  &   $4.22\pm0.52$  &   $6.47\pm0.37$  &   $0.606\pm0.01$  &  $0.36\pm0.01$  &    8.00 &   0.023 &   0.997 \\
       240 & $354\pm8$  &  $1.09\pm0.14$  &   $918\pm19$  &   $4.38\pm0.60$  &   $6.50\pm0.39$  &   $0.577\pm0.01$  &  $0.36\pm0.02$  &    8.19 &   0.023 &   0.996 \\
\hline
\end{tabular}  
}
\end{table*}

\begin{figure}
\begin{center}
\includegraphics[scale=0.50, angle=0]{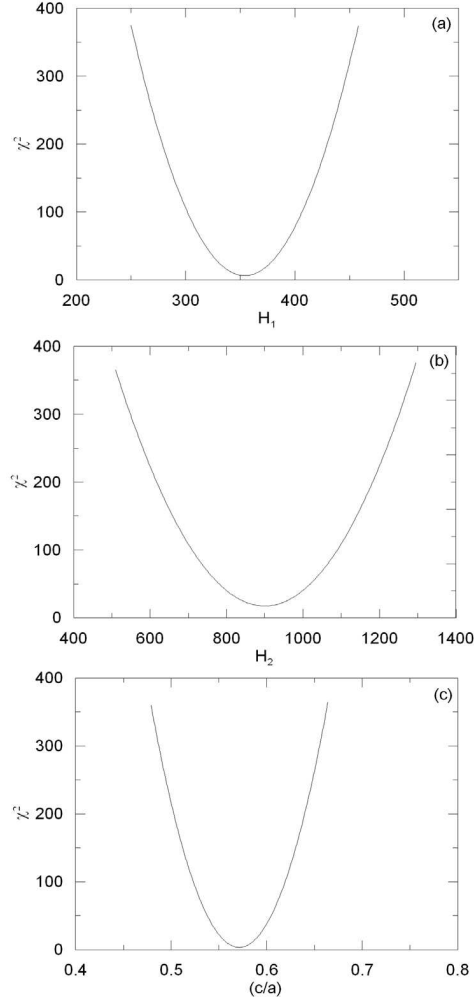}
\caption[]{$\chi^{2}$ distribution of estimated $H_{1}$ (a), $H_{2}$ (b) and $(c/a)$ (c) model parameters for the star field with $l=60^{\circ}$.}
\label{Fig07}
\end{center}
\end {figure}

\begin{figure}
\begin{center}
\includegraphics[scale=0.40, angle=0]{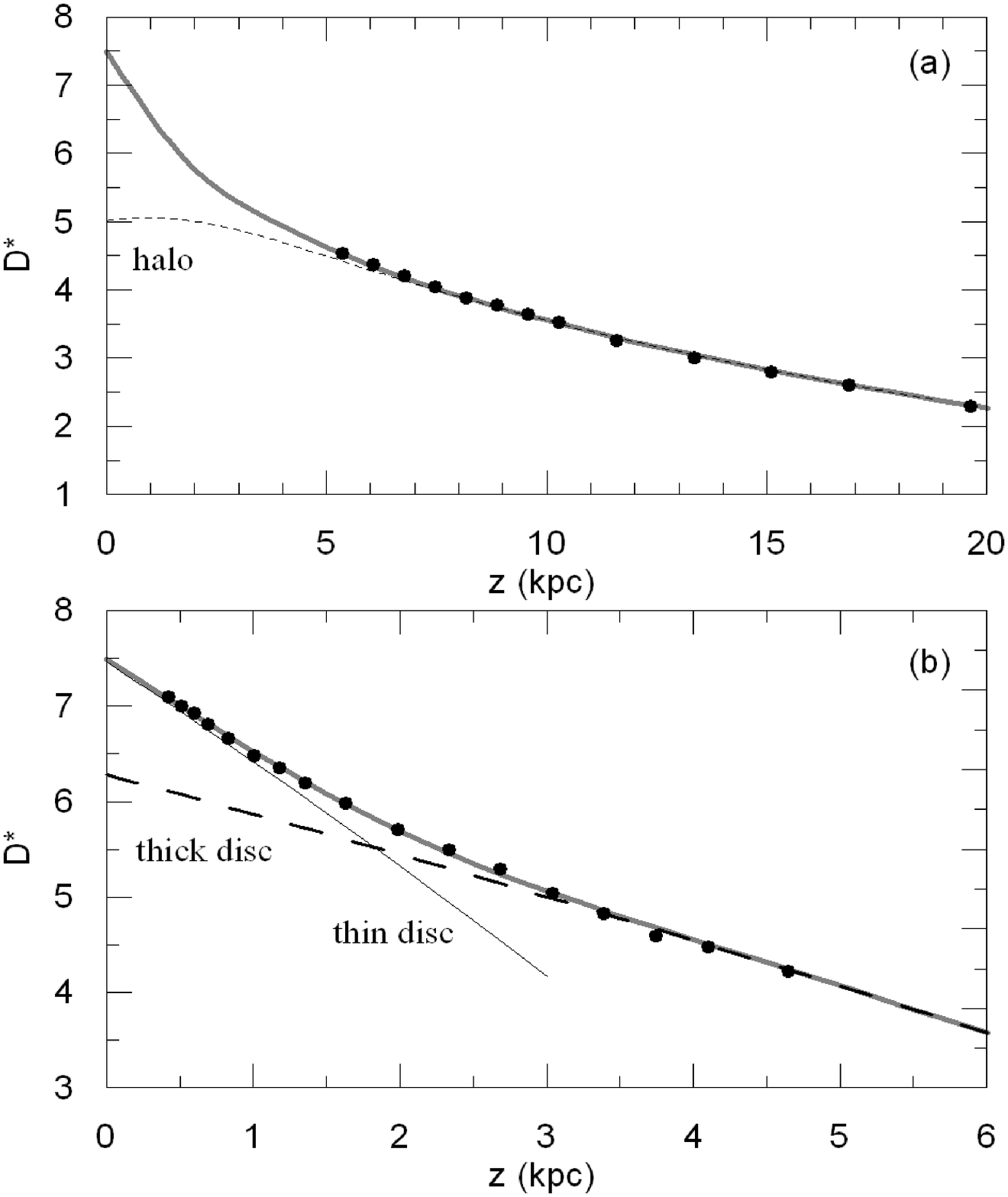}
\caption[]{Investigation for degeneracy in model parameters of simultaneously estimated populations for the star field with $l=60^{\circ}$. 
The dots and thick line represent observed and theoretical space densities, respectively. (a) Using $z>5$ kpc data, only the local 
space density and the axial ratio of the halo are estimated independent of discs. (b) Data for $z\leq 5$ kpc without halo 
densities are used to estimate the model parameters of thin and thick discs.} 
\label{Fig08}
\end{center}
\end {figure}

We also used a different procedure to test any possible degeneracy in the estimation of the model parameters. 
First, we estimated the local space density and the axial ratio for the halo by comparing the logarithmic space 
density function for $z>5$ kpc with the analytical density law of the halo (Eq. 16). Then, we omitted the space 
density of the halo, estimated by the corresponding density law, and compared the new density function for $z\leq5$ 
kpc with the combined density laws of thin and thick discs (Fig. 8b). This procedure provides galactic model 
parameters for thin and thick discs independent of the model parameters of the halo. The result of the application 
of two different procedures shows that the corresponding galactic model parameters for a specific population 
are either identical or differ only by a negligible amount. Hence, we may argue that there is no degeneracy 
in the estimated parameters.

A similar procedure is applied to the thin and thick discs. We estimated the local space densities and the scaleheights 
of thin and thick discs simultaneously by using the space density function for $0.3<z\leq5$ kpc and compared the model parameters 
of the thick disc with the corresponding ones, which was estimated by using space density function for $1.5<z\leq5$ kpc, where the 
thick disc is dominant. Since no significant differences could be observed between the compared parameters, we concluded 
that no degeneracy exists between the two discs either.

\subsection{The Trend of the Local Space Density of the Thick Disc and Halo}

The local space densities of the thick disc ($n_{2}/n_{1}$) and halo ($n_{3}/n_{1}$), relative to the local space density 
of the thin disc, were plotted versus the galactic longitude in Fig. 9. The trend is flat in both panels with exception of 
two small offsets at $l=60^{\circ}$ and $l=210^{\circ}$ in panel (b). This invariance against galactic longitude is not 
mentioned in the literature, therefore it is new. However, it is different than the results obtained for high latitude 
fields. For example, in our recent works (\citealt{Karaali07, Bilir08a}) the local space densities of thick disc ($n_{2}/n_{1}$) and 
halo ($n_{3}/n_{1}$) vary with galactic longitude (see Section 4).

\begin{figure}
\begin{center}
\includegraphics[scale=0.40, angle=0]{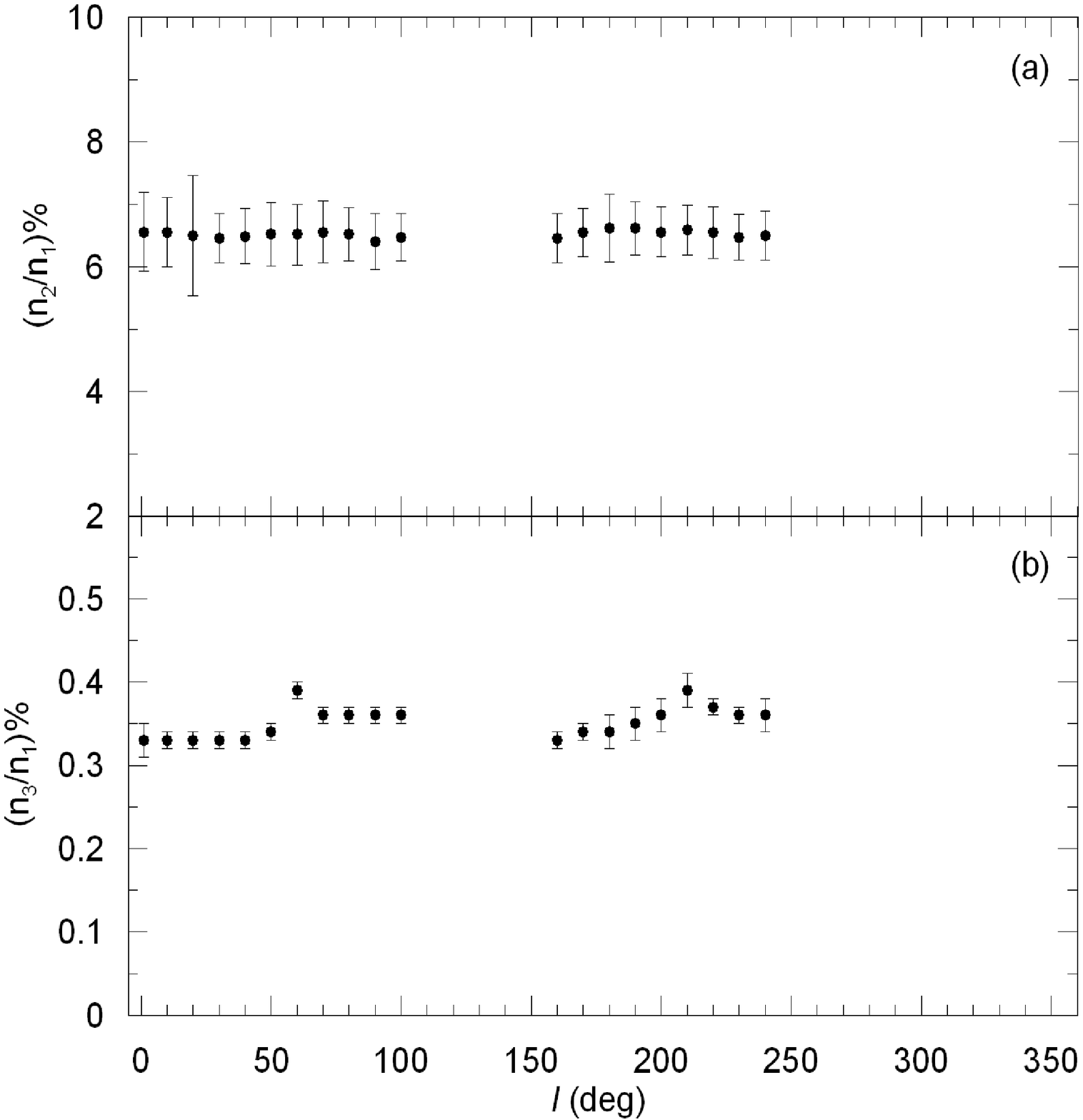}
\caption[]{Variation of normalized local space densities of thick disc (a) and halo (b) as a function of galactic longitude.} 
\label{Fig09}
\end{center}
\end {figure}

\subsection{Dependence of the Scaleheights and Scalelengths of the Discs and Axial Ratio of the Halo on the Galactic Longitude}

The variation of the scaleheights of the thin and thick discs, $H_{1}$ and $H_{2}$, respectively, with the galactic 
longitude ($l$) were plotted in Fig. 10. $H_{1}$ assumes its minimum (325 pc) in the galactic centre direction and 
it increases with longitude in the first quadrant. In the second quadrant, it gives the indication of decreasing, 
however it is not complete. Finally, the third quadrant displays a slight variation of $H_{1}$ with $l$, where $H_{1}$ 
assumes its maximum (369 pc).

\begin{figure}
\begin{center}
\includegraphics[scale=0.40, angle=0]{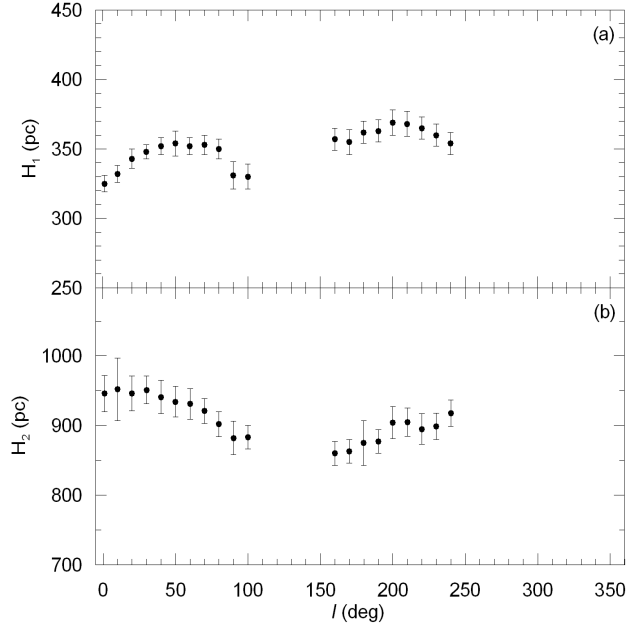}
\caption[]{Variation of the scaleheights of the thin (a) and thick discs (b) as a function of galactic longitude.} 
\label{Fig10}
\end{center}
\end {figure}

The variance of thick disc's scaleheight ($H_{2}$) with galactic longitude differs from variance of thin disc's scaleheight ($H_{1}$). 
$H_{2}$ assumes its maximum (952 pc) at the galactic longitude $l=20^{\circ}$ and it decreases within the first quadrant. It assumes 
the minimum value 860 pc at $l=160^{\circ}$ and it increases gradually in the third quadrant. One can see a similarity in the trends of 
scaleheight of the thick disc ($H_{2}$) and surface densities of the fields investigated (Fig. 1).

The numerical values as well as the trends of the scalelengths of two discs are also different (Fig. 11). The scalelength 
of the thin disc ($h_{1}$) is almost constant ($\approx$ 1 kpc) in the longitude interval $0^{\circ}\leq l \leq60^{\circ}$, 
then it increases with high errors up to $l=100^{\circ}$. The trend of $H_{1}$ in the third quadrant resembles a parabolic 
function with a maximum of 1.69 kpc at $l=200^{\circ}$.

\begin{figure}
\begin{center}
\includegraphics[scale=0.45, angle=0]{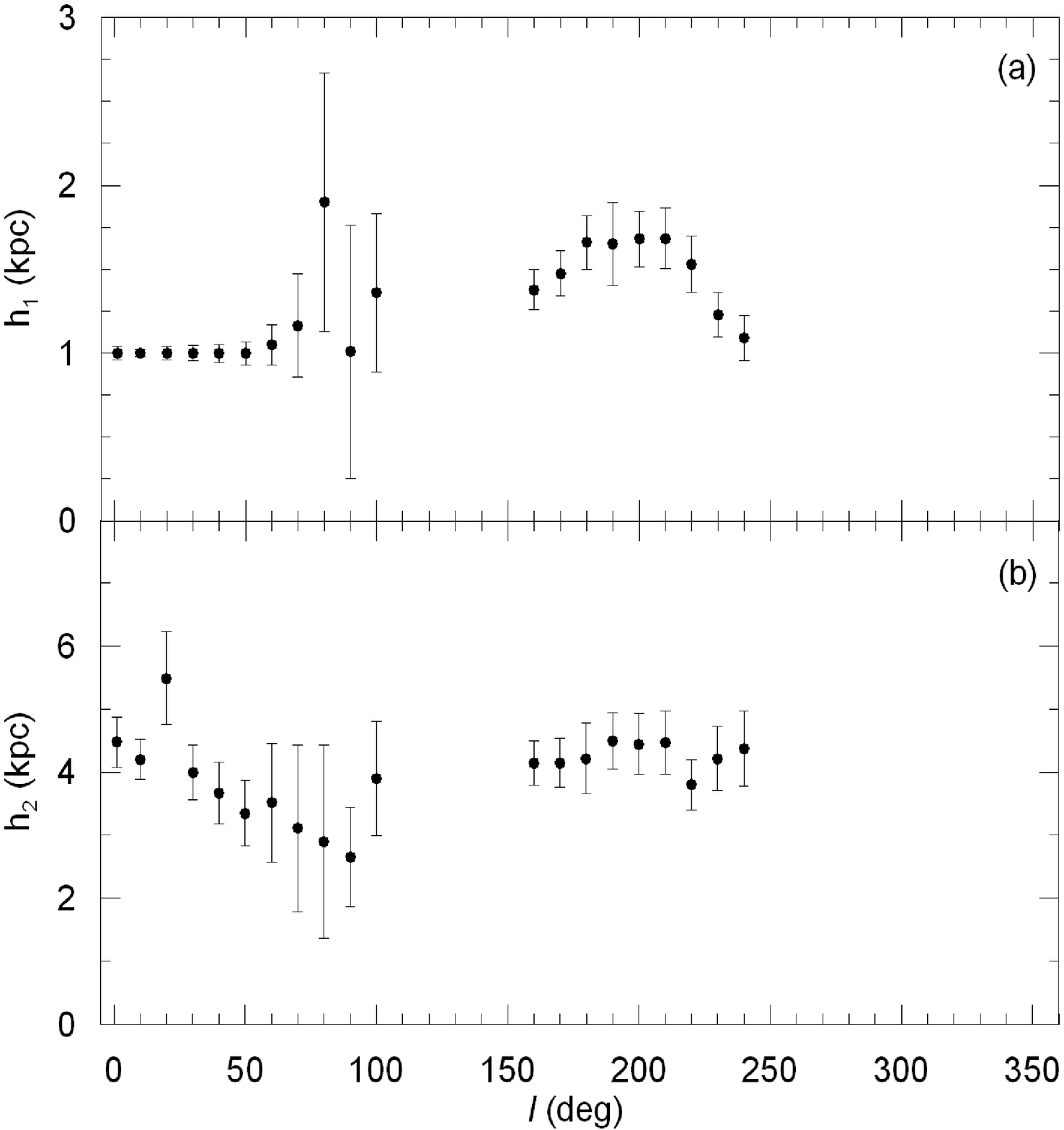}
\caption[]{Variation of scalelengths of the thin (a) and thick (b) discs as a function of galactic longitude.} 
\label{Fig11}
\end{center}
\end {figure}

The scalelength of the thick disc ($h_{2}$) decreases with galactic longitude in the first quadrant and it assumes 
its maximum and minimum, 5.49 and 2.65 kpc at $l=20^{\circ}$ and $l=90^{\circ}$, respectively. However, it is 
rather flat in the third quadrant ($h_{2}\approx4.26$ kpc).   

The variation of the axial ratio of the halo ($c/a$) is given in Fig. 12. The trend of ($c/a$) gives the indication of 
a smooth variation ($0.55 \leq (c/a) \leq 0.60$) except, perhaps, a slight minimum in the second quadrant and a small deviation 
towards large values of ($c/a$), between galactic longitudes $l=200^{\circ}$ and $l=230^{\circ}$. That is, the halo component of 
the Galaxy has a disclike structure at the intermediate galactic latitudes which is what we were expecting (see Section 4).

\begin{figure}
\begin{center}
\includegraphics[scale=0.40, angle=0]{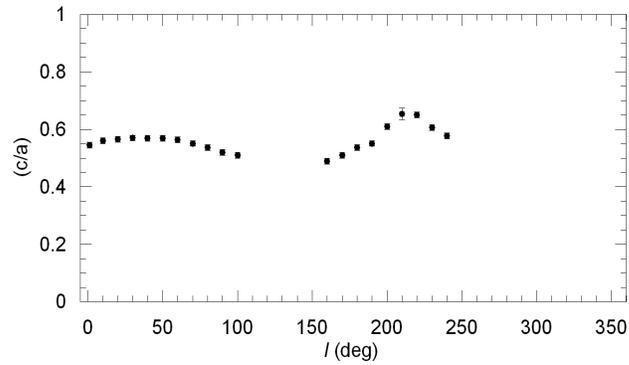}
\caption[]{Variation of axial ratio of the halo as a function of galactic longitude.} 
\label{Fig12}
\end{center}
\end {figure}

\subsection{Effect of the Unsolved Binaries}
A high fraction of stars are actually binary systems and being a binary system makes stars appear brighter 
and redder than normally they are. Different fractional values (defined as $f$) can be found in the literature. 
For example, using the data in the Gliese catalogue of nearby stars, \cite{Brosche64} found a value of $f=0.4$ 
due to his simple model for the resolution criterion. On considering the local (within a distance of 10 pc) 
binary fraction, \cite{Reid91} concluded that the proportion of binaries among ``stars'' is consistent with a 
value ranging from 30 to 50\%. When all systems in question are binary stars, i.e. $f=1$, \cite{Kroupa91} 
found that a single mass function provides the best representation of a single luminosity function. However, 
a smaller value can not be discarded with high confidence. \cite{Halbwachs86} used all available data on binary systems 
and concluded that the proportion of single stars among all stellar systems is at most 23\% when spectroscopic 
binaries are taken into account. An extensive long  term radial velocity study of the Hyades cluster reveals that 
at least 30\% of the cluster stars are spectroscopic binaries and that essentially all stars brighter than 
the Hyades mainsequence stars are actually binary systems \citep{Griffin88}.

The effect of binary stars were discussed in (\citealt{Kroupa90,Kroupa93}, hereafter KTG90 and KTG93, respectively) 
extensively as well as other effects such as metallicity, age, distance etc. KTG93 adopt the 
binary fraction $f\sim0.6-0.7$ as a reasonable value. They give the mentioned combined effects as ``cosmic scatter'' 
as a function of $(V-I)$ colour in the range $0.5<(V-I)<4.5$. These authors estimate the scatter belonging 
to binaries alone as $\sigma=0.27$ mag, if a fraction $f=0.8$ of all stars are unresolved binary systems.

The effect of binary stars on the Galactic model parameters were discussed by many authors 
(cf. \citealt{Siegel02, J08, Ivezic08}). The net effect of binarism, stated in the literature, is an underestimation 
of the scaleheights for thin and thick discs. We adopted three fractions of binary stars, i.e. 25\%, 50\% and 75\%, 
and evaluated all the Galactic model parameters for thin and thick discs, and halo for one of our fields ($l =60^{\circ}$). 
The procedure is similar to that of \cite{Siegel02}. For a fraction of stars, we added a companion chosen at random 
to be equal in mass to the primary. The colour and magnitude of each primary star was then reset to the combined 
characteristics of the binary. Then, we evaluated the density function of the stars using the original calibrations 
and inferred density law was compared with the input value (Fig. 13). The result is a steepening of the measured 
density law and consequent underestimation of the scaleheight. Whereas effect of binarism causes an overestimation for the 
local space densities (Table 2).

According to our results, a fraction of 50\% unresolved binaries decreases the scaleheight of the thin 
disc by 22\%, for a specific field ($l =60^{\circ}$), which is between the values stated by \cite{J08} 
and \cite{Siegel02}, i.e. 20 and 25\%, respectively. However, the decrease of the scaleheight for the 
thick disc due to binarism is higher, 45\%, than the one stated by \cite{Siegel02}, 20-29\%. 

\begin{figure}
\begin{center}
\includegraphics[scale=0.40, angle=0]{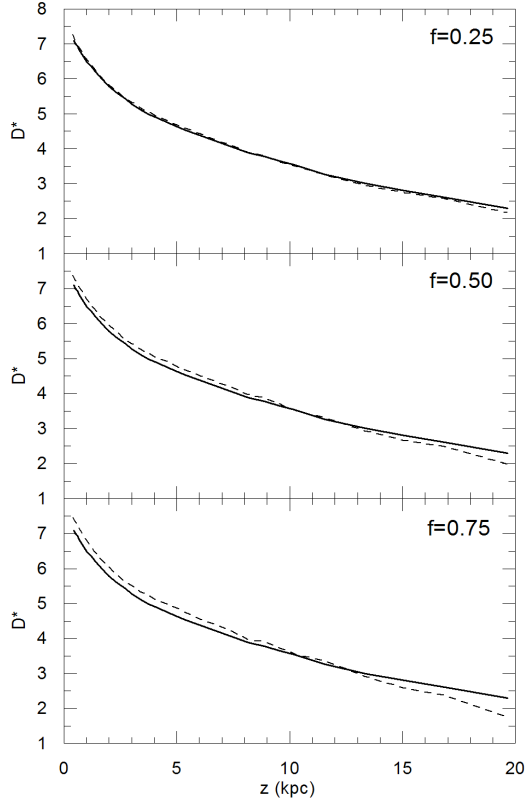}
\caption[]{The effect of unresolved binaries upon the derived density law for different ratio of equal mass binaries. 
The panels show the actual density law (solid line) against the derived density law (dashed line) for various binary fraction.}
\label{Fig13} 
\end{center}
\end{figure}

\begin{table}
\setlength{\tabcolsep}{1.5pt}
{\tiny
\center
\caption{The effect of unresolved binaries upon the Galactic model parameters for thin and thick discs, and halo, 
for three fractions of binary stars for the field with $l=60^{\circ}$ (distances in pc).}
\begin{tabular}{rccccccc}
\hline
                         &     &\multicolumn{2}{c}{f=0.25}            &   \multicolumn{2}{c}{f=0.50} &          \multicolumn{2}{c}{f=0.75}\\
\hline
 Parameter & Actual Value & Derieved Value & Variation ($\%$) & Derieved Value & Variation ($\%$) & Derieved Value & Variation ($\%$) \\
\hline
         $H_{1}$ (pc) &       352 &        376      &          7       &        428     &         22       &        464     &         32 \\
         $H_{2}$ (pc) &       931 &       1069      &         15       &       1350     &         45       &       1465     &         57 \\
         (c/a)        &      0.56 &       0.55      &          2       &       0.53     &          5       &       0.51     &          9 \\
$n_{2}/n_{1}$ ($\%$)  &      6.52 &       5.51      &         15       &        4.3     &         34       &        3.9     &         40 \\
$n_{3}/n_{1}$ ($\%$)  &      0.39 &       0.38      &          3       &       0.37     &          5       &       0.36     &          8 \\
\hline
\end{tabular}
} 
\end{table}

\subsection{Metallicity Distribution}

The photometric metallicities of about 130 000 G-type stars with absolute magnitudes $5 < M_{g} \leq 6$ were evaluated 
using the following equation of KBT:

\begin{eqnarray}
[M/H]= 0.10-3.54\delta_{0.43}-39.63\delta_{0.43}^{2}+63.51\delta_{0.43}^{3}.
\end{eqnarray}
 
This equation was calibrated for the main-sequence stars with $0.12<(g - r)_{0} \leq0.95$ which covers the $(g - r)_{0}$ 
colour indices of our sample. Here, $\delta_{0.43}$ is the normalized UV-excess in {\em SDSS} photometry corresponding to 
$\delta_{0.6}$ in the UBV photometry. KBT give the range of the metallicity as $-2.76\leq [M/H] \leq0.20$ dex, 
corresponding to $0< \delta_{0.43} \leq0.33$.

The metallicity distribution for each field is determined using a mean metal-abundance value determined for each of the following 
distance intervals (in kpc): (0.5, 1.5], (1.5, 2], (2, 2.5], (2.5, 3], (3, 4], (4, 5], (5, 6], (6, 7], (7, 8], (8, 9], 
(9, 12], (12, 15]. Fig. 14 shows the metallicity distribution for the field with galactic longitude $l=60^{\circ}$, as an 
example. The projection of the centroid distances ($r^{*}$) of these intervals onto the vertical direction, i.e. 
$z^{*}=r^{*}\sin b$, are as follows: 0.85, 1.26, 1.61, 1.96, 2.52, 3.22, 3.92, 4.62, 5.33, 6.03, 7.57, 9.66 kpc. 
The mean metal abundance is adopted as the mode of the Gaussian curve fitted to each distribution, except three intervals, 
(4,5], (5,6] and (6,7], where the metallicity distribution is almost flat, and where the median was assumed as the mean of the 
metallicity distribution. The variation of the mean metallicities with $z^{*}$ distances for 20 fields is given in Table 3. 
The metallicity histogram, obtained from Table 3, for all fields is given in Fig. 15 for each $z^{*}$ distance. 

\begin{figure}
\begin{center}
\includegraphics[scale=0.70, angle=0]{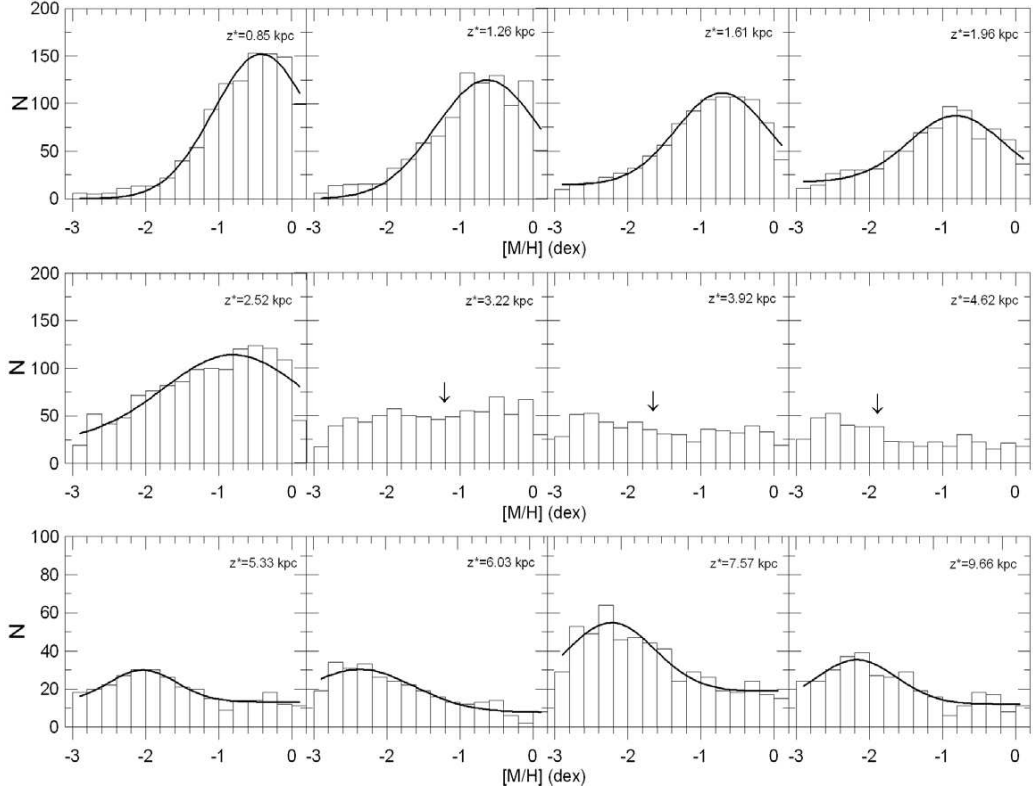}
\caption[]{Metallicity distributions as a function of vertical distance $z^{*}$ for the field with $l=60^{\circ}$.}
\label{Fig14} 
\end{center}
\end{figure}

\begin{figure}
\begin{center}
\includegraphics[scale=0.70, angle=0]{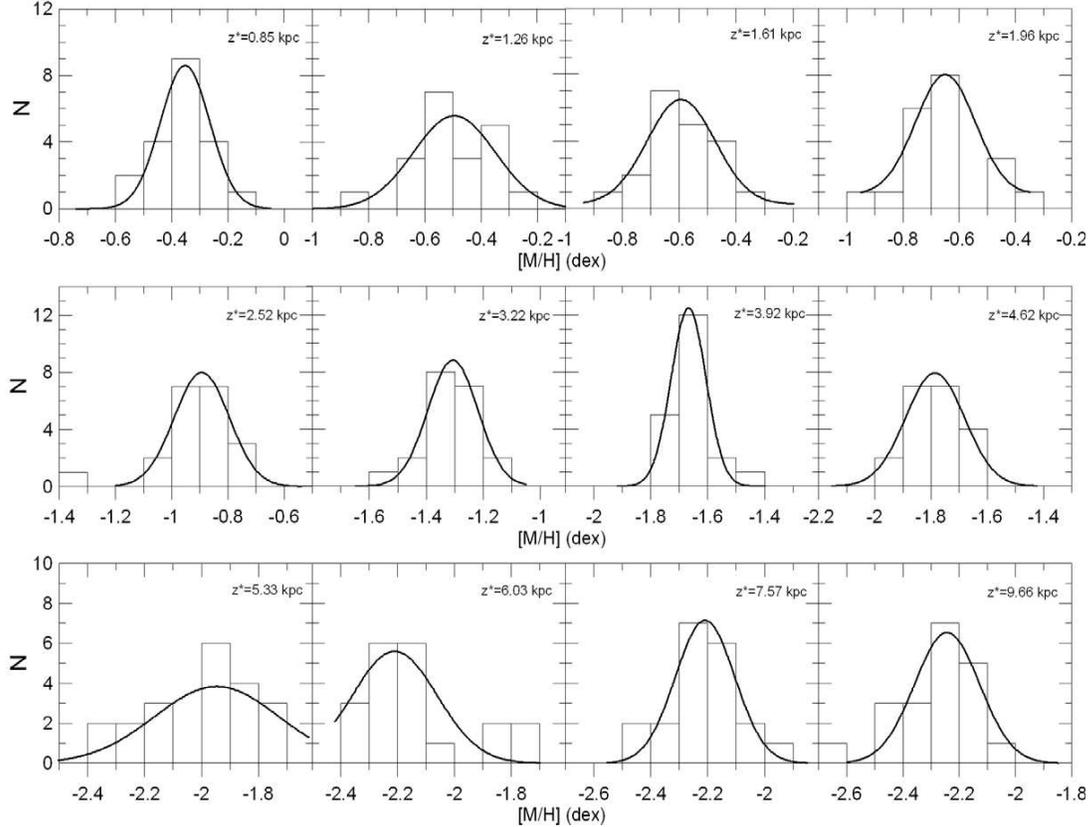}
\caption[]{Variation of the metallicities, for 20 star fields, with vertical distance $z^{*}$.}
\label{Fig15} 
\end{center}
\end{figure}

We coupled the mode/median and the corresponding $z^{*}$ distance for each histogram (Table 4) and plotted them in Fig. 16 
in order to investigate the vertical metallicity gradient for 20 intermediate latitude fields. One can see four different 
trends in the figure: (1) for relatively short $z$ distances, i.e. $z^{*}<2.5$ kpc, the variation of $[M/H]$ is rather smooth; 
(2) for intermediate $z^{*}$ distances, i.e. $2.5 \leq z^{*}< 4$ kpc, the variation is steeper but still smooth; (3) for 
$3.5 \leq z^{*}< 5.5$ kpc, the variation is the same as for the short distances; (4) and finally for $6 < z^{*}< 10$ kpc 
there is no variation. The metallicity gradient describing the first trend, $d[M/H]/dz=-0.32$ dex kpc$^{-1}$, is in agreement 
with the canonical metallicity gradients for the same $z^{*}$ distances, and is likely the signature of this galactic regions 
formation by a dissipative collapse. The description of the second trend in terms of a metallicity gradient, 
$d[M/H]/dz=-0.56$ dex kpc$^{-1}$, corresponds to the average metallicity differences between the three population components 
involved. The third trend, $d[M/H]/dz=-0.20$ dex kpc$^{-1}$, corresponds to the metallicity difference between thick disc 
and inner halo, and finally the fourth trend, $d[M/H]/dz=-0.01$ dex kpc$^{-1}$ is the very low metallicity gradient of the 
inner spheroid. The mean metallicity gradient for the whole $z^{*}$ interval, $z^{*}<10$ kpc, $d[M/H]/dz=-0.25$ dex kpc$^{-1}$, 
is in agreement with the ones appearing in the literature for high latitude fields \citep{Ak07a, Ak07b}. 

\begin{figure}
\begin{center}
\includegraphics[scale=0.50, angle=0]{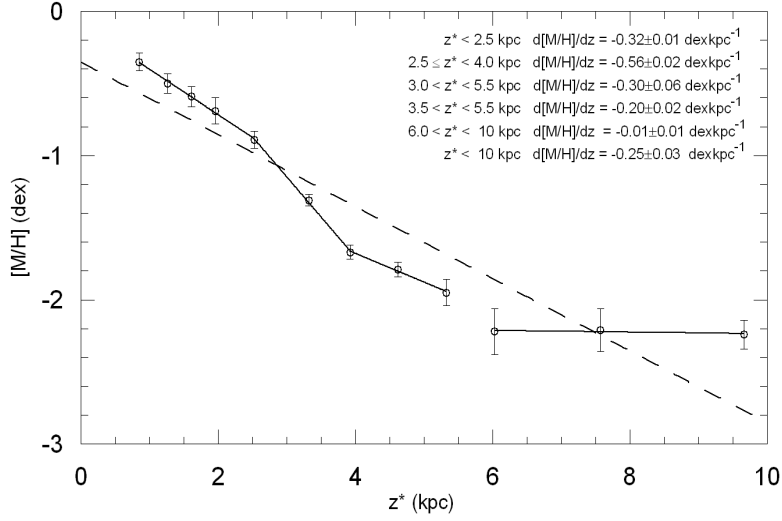}
\caption[]{Metallicity versus vertical distance $z^{*}$. \bf{The dashed line represents the mean metallicity gradient.}}
\label{Fig16} 
\end{center}
\end{figure}

\begin{table*}
\setlength{\tabcolsep}{2pt}
{\tiny
\center
\caption{The metallicity distribution as a function of distance to the galactic plane, for 20 intermediate latitude ($\langle b \rangle=45^\circ$) fields at different galactic longitudes ($0^{\circ} \leq l \leq100^{\circ}$ and $160^{\circ}\leq l \leq240^{\circ}$). Uncertainties (given in parentheses) refer to the last quoted digits. Distances are in kpc.}
\begin{tabular}{cccccccccccccc}
\hline
 $l/z^{*}$ &       0.85 &       1.26 &       1.61 &       1.96 &       2.52 &       3.32 &       3.92 &       4.62 &       5.33 &       6.03 &       7.57 &       9.66 &       N \\
   deg/kpc &            &            &            &            &            &            &            &            &            &            &            &            &         \\
\hline
         0 &  -0.58(06) &  -0.82(05) &  -0.90(05) &  -0.97(07) &  -1.35(02) &  -1.55(02) &  -1.62(02) &  -1.61(02) &  -2.34(09) &  -2.19(19) &  -2.27(16) &  -2.14(23) &   10058 \\
        10 &  -0.34(03) &  -0.58(05) &  -0.68(04) &  -0.70(07) &  -0.83(14) &  -1.50(02) &  -1.64(03) &  -1.80(03) &  -1.95(19) &  -2.20(09) &  -1.98(15) &  -2.34(05) &   13623 \\
        20 &  -0.28(06) &  -0.49(09) &  -0.52(05) &  -0.62(05) &  -0.82(19) &  -1.30(02) &  -1.60(03) &  -1.76(03) &  -1.85(09) &  -1.78(13) &  -2.24(09) &  -2.24(11) &   13607 \\
        30 &  -0.41(05) &  -0.59(04) &  -0.65(07) &  -0.69(05) &  -0.80(02) &  -1.37(02) &  -1.69(03) &  -1.68(03) &  -2.17(08) &  -2.22(29) &  -2.04(09) &  -2.40(02) &   12765 \\
        40 &  -0.54(05) &  -0.65(04) &  -0.75(04) &  -0.78(06) &  -0.81(07) &  -1.44(03) &  -1.70(03) &  -1.76(03) &  -1.74(04) &  -2.09(17) &  -2.15(08) &  -2.36(10) &   11771 \\
        50 &  -0.46(05) &  -0.61(05) &  -0.70(05) &  -0.66(05) &  -0.84(09) &  -1.32(03) &  -1.57(03) &  -1.84(04) &  -2.28(09) &  -2.30(10) &  -2.41(43) &  -2.41(09) &   10368 \\
        60 &  -0.42(03) &  -0.63(06) &  -0.71(03) &  -0.82(05) &  -0.80(11) &  -1.24(03) &  -1.63(04) &  -1.89(04) &  -2.00(20) &  -2.37(12) &  -2.23(07) &  -2.17(07) &    8430 \\
        70 &  -0.47(05) &  -0.52(06) &  -0.61(03) &  -0.75(05) &  -1.00(03) &  -1.37(04) &  -1.64(04) &  -1.74(05) &  -1.77(05) &  -1.76(06) &  -2.27(08) &  -2.10(03) &    7101 \\
        80 &  -0.38(04) &  -0.56(05) &  -0.63(04) &  -0.77(04) &  -0.99(03) &  -1.33(04) &  -1.71(05) &  -1.77(05) &  -2.16(06) &  -2.17(25) &  -2.40(16) &  -2.30(14) &    5648 \\
        90 &  -0.34(06) &  -0.29(05) &  -0.44(06) &  -0.49(16) &  -0.90(03) &  -1.23(04) &  -1.50(05) &  -1.70(06) &  -1.86(07) &  -2.24(25) &  -2.36(24) &  -2.27(13) &    4102 \\
       100 &  -0.32(08) &  -0.51(05) &  -0.59(07) &  -0.74(07) &  -0.79(08) &  -1.30(05) &  -1.66(06) &  -1.89(06) &  -2.00(07) &  -2.21(13) &  -2.10(17) &  -2.30(04) &    3635 \\
       160 &  -0.34(07) &  -0.36(12) &  -0.66(12) &  -0.71(12) &  -1.00(04) &  -1.30(06) &  -1.73(07) &  -1.70(07) &  -1.80(07) &  -1.84(19) &  -2.13(19) &  -2.20(17) &    2701 \\
       170 &  -0.32(05) &  -0.40(08) &  -0.49(07) &  -0.46(16) &  -0.94(05) &  -1.37(06) &  -1.65(07) &  -1.88(07) &  -2.00(06) &  -2.18(05) &  -2.20(04) &  -2.19(10) &    2492 \\
       180 &  -0.38(06) &  -0.43(10) &  -0.42(11) &  -0.64(09) &  -0.99(05) &  -1.33(06) &  -1.70(07) &  -1.80(05) &  -1.86(08) &  -2.19(17) &  -2.16(27) &  -2.30(06) &    2704 \\
       190 &  -0.26(09) &  -0.38(05) &  -0.44(20) &  -0.44(09) &  -0.88(05) &  -1.30(06) &  -1.73(06) &  -1.88(07) &  -2.23(09) &  -2.20(33) &  -2.27(28) &  -2.24(21) &    2761 \\
       200 &  -0.18(03) &  -0.32(11) &  -0.39(03) &  -0.39(23) &  -0.86(05) &  -1.20(05) &  -1.74(06) &  -1.80(06) &  -1.83(07) &  -1.90(12) &  -2.18(09) &  -2.43(11) &    3062 \\
       210 &  -0.30(07) &  -0.48(07) &  -0.60(07) &  -0.62(10) &  -1.05(04) &  -1.31(05) &  -1.79(06) &  -1.96(06) &  -2.20(04) &  -2.24(15) &  -2.29(14) &  -2.41(23) &    3684 \\
       220 &  -0.37(07) &  -0.35(09) &  -0.58(06) &  -0.61(08) &  -0.98(04) &  -1.17(06) &  -1.61(06) &  -1.87(07) &  -1.98(06) &  -2.40(07) &  -2.20(03) &  -2.20(04) &    3570 \\
       230 &  -0.24(08) &  -0.54(03) &  -0.64(05) &  -0.77(03) &  -0.97(04) &  -1.30(05) &  -1.67(06) &  -1.85(06) &  -2.31(31) &  -2.31(12) &  -2.49(18) &  -2.69(06) &    3526 \\
       240 &  -0.33(06) &  -0.52(06) &  -0.58(06) &  -0.61(08) &  -1.04(04) &  -1.35(05) &  -1.70(05) &  -1.93(06) &  -2.00(05) &  -2.22(17) &  -2.28(08) &  -2.30(03) &    3761 \\
\hline
\end{tabular}  
}
\end{table*}

\begin{table}
\center
\caption{Mean metallicities (for different distance intervals) calculated from the metallicity distributions of 20 fields. 
Symbols: $r$: distance from the Sun; $r^{*}$: centroid distance corresponding to the interval $r_{1}-r_{2}$; $z$: projection 
of $r^{*}$ onto the vertical direction; $[M/H]$: mean metallicity.}
\begin{tabular}{cccc}
\hline
  $r_{1}-r_{2}$ &  $r^{*}$   &     $z^{*}$&      [M/H] \\
     (kpc)      &      (kpc) &      (kpc) &      (dex) \\
\hline
   0.5-1.5      &       1.21 &       0.85 &  -0.35(06) \\
   1.5-2.0      &       1.79 &       1.26 &  -0.50(07) \\
   2.0-2.5      &       2.28 &       1.61 &  -0.59(07) \\
   2.5-3.0      &       2.77 &       1.96 &  -0.69(09) \\
   3.0-4.0      &       3.57 &       2.52 &  -0.89(06) \\
   4.0-5.0      &       4.55 &       3.32 &  -1.31(04) \\
   5.0-6.0      &       5.55 &       3.92 &  -1.67(05) \\
   6.0-7.0      &       6.54 &       4.62 &  -1.79(05) \\
   7.0-8.0      &       7.53 &       5.33 &  -1.95(09) \\
   8.0-9.0      &       8.53 &       6.03 &  -2.22(16) \\
  9.0-12.0      &      10.71 &       7.57 &  -2.21(15) \\
 12.0-15.0      &      13.66 &       9.66 &  -2.24(10) \\
\hline
\end{tabular}  
\end{table}

\section{Summary and Discussion}

We estimated the galactic model parameters for 20 intermediate-latitude fields with galactic longitudes $0^{\circ}\leq l \leq 100^{\circ}$ 
and $160^{\circ}\leq l \leq 240^{\circ}$ to explore their possible variation with galactic longitude. We evaluated the metal 
abundance for about 130 000 G-type main-sequence stars with absolute magnitudes $5<M_{g} \leq6$ to investigate the metallicity gradient of 
different components of the Galaxy, as well.

\subsection{Local Space Densities}

The local space densities of the thick disc and halo relative to the local space densities of the thin disc 
with galactic longitude do not vary. This is an innovation and it confirms its definition. However its trend is different than the 
results obtained for high latitude fields \citep{Karaali07, Bilir08a}. The dependence of the local space densities of thick 
disc and halo on galactic latitude for high latitude fields is a result of a bias effect. High latitudes give the chance 
to the investigator to reach greater distances in the vertical direction, for a specific distance in the line of sight. 
That is, in high latitudes, the halo stars are dominant and they affect the galactic model as their contribution results in 
different local space densities, depending on the number of halo stars.       

\subsection{Scaleheight and Scalelength of the Thin Disc}

As cited in Section 3.3, the scaleheight and scalelength of the thin disc vary with galactic longitude, however, with different 
trends. These variations are explained as a disc-flare. \cite{Lopez02} showed that the scaleheight of the thin 
disc increases with galactocentric distance ($R$):

\begin{eqnarray}
h_{z}=h_{z,o}[1+0.21(R-R_{0})+0.056(R-R_{0})^{2}], 
\end{eqnarray}

where $R_{0}$ is the galactocentric distance of the Sun and $h_{z,o}$ is the scaleheight of the thin disc at the solar 
distance. We plotted the scaleheights of the thin disc for 20 fields versus galactocentric distances in order to confirm 
this equation. Fig. 17 shows that the scaleheight of the thin disc increases with galactocentric distance and supplies 
the expected confirmation:
\begin{eqnarray}	
H_{1}=(349.68\pm 39.91)[1+(0.053\pm 0.013)(R-R_{0}).
\end{eqnarray}

\begin{figure}
\begin{center}
\includegraphics[angle=0, width=80mm, height=94.1mm]{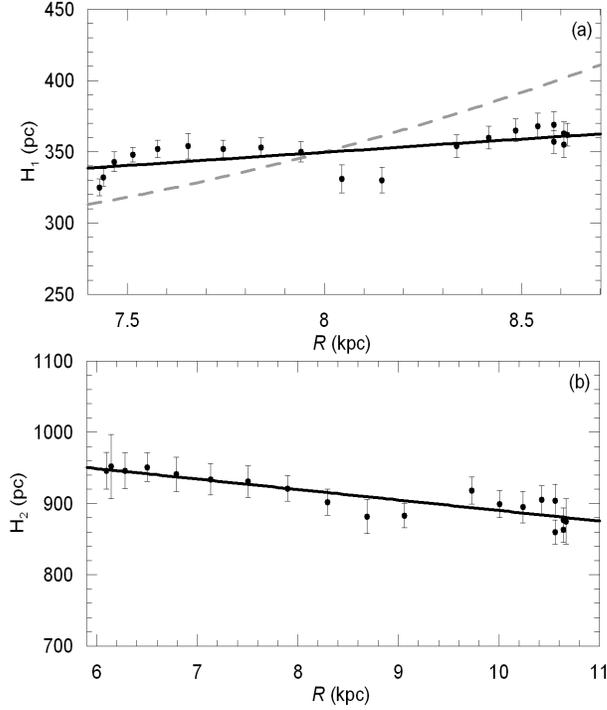}
\caption[]{Variation of the scaleheight of the thin (a) and thick (b) discs with $R$. The dots, the dashed and solid lines 
represent estimated scaleheights, using adopting \citet{Lopez02}'s flare model (Eq. 21) and calculated flare model using 
scaleheight of 20 star fields (Eq. 22).}
\label{Fig17}
\end{center}
\end {figure}

\subsection{Scaleheight and Scalelength of the Thick Disc}

The variation of the scaleheight and scalelength of the thick disc with galactic longitude are due to: 1) the long bar, and 
2) the flares. The bar induces a gravitational ``wake'', traps and piles up stars behind it (\citealt{Hernquist92, Debattista98}). 
Thus, in response to a bar, one expects an excess of stars in the direction of the long bar ($l\sim 27^{\circ}$ and 
$\sim 207^{\circ}$) and a deficiency of stars in the opposite direction, which affects the numerical values of the scaleheight 
and scalelength of the thick disc. It is worth noting that the maximum surface density for stars with apparent magnitudes 
$15 < g_{0} \leq 18$ (where thick disc stars are dominated) corresponds to galactic longitude $\sim 20^{\circ}$ (Fig. 18).

\begin{figure}
\begin{center}
\includegraphics[scale=0.50, angle=0]{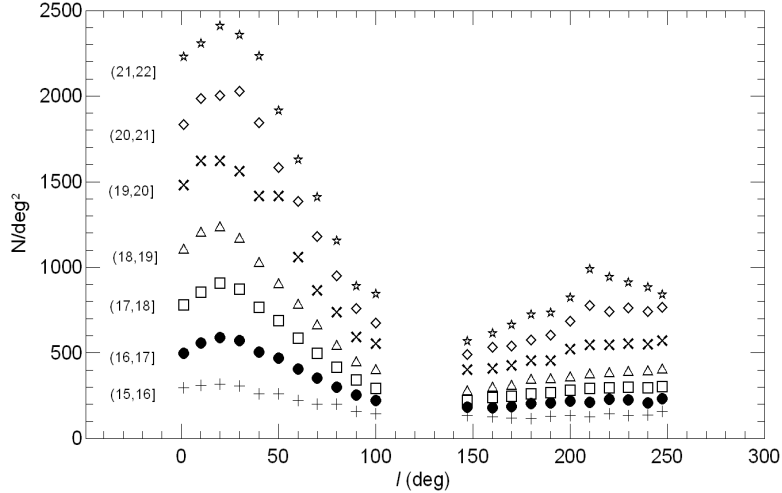}
\caption[]{Variation of the number of stars with $g_{0}$ apparent magnitude intervals as a function of the galactic longitude.} 
\label{Fig18}
\end{center}
\end{figure}

Also, the flare produces a change in the scaleheight of the thick disc as the galactocentric distance varies. 
This effect is well explained in \citet{Lopez02, Lopez04} and \citet{Cabrera-Lavers07}, who showed that the thick disc 
presents a flaring in the opposite sense of the thin disc. That is, the scaleheight of the thick disc decreases with  
increasing galactocentric distance \citep{Lopez02, Momany06}. We plotted the scaleheights versus the galactocentric 
distances of 20 fields and confirmed this suggestion. Actually, panel (b) in Fig. 17 shows that the scaleheight of the 
thick disc decreases with galactocentric distance. We are quoting the work of \citet{Bilir08a} where the same 
confirmation was done.

\subsection{Axial Ratio of the Halo}

The trend of the axial ratio $(c/a)$ of the halo with galactic longitude is almost flat ($\langle c/a \rangle$=0.56), i.e. its range for  
75\% of the fields is only $10^{\circ}$. However, there is a slight minimum and a small maximum corresponding to the fields 
in the second and third quadrants, respectively. The flatness of $(c/a)$ can be exposed by the fact that the halo 
component of the Galaxy investigated in this work corresponds to the ``inner halo''. Since the maximum galactocentric distance of the 
fields is less than the upper limit of the inner halo, i.e: $R$:10-15 kpc \citep{Corollo07}, this statement is confirmed. Also, the inner 
halo has a dislike structure which is confirmed by the mean axial ratio ($\langle c/a \rangle$=0.56) estimated in our work. Additionally, 
the inner halo is the outermost part of the Galaxy which can be investigated by the use of intermediate galactic fields. 

The variation of the halo model parameters with galactic latitude and longitude estimated in high latitude fields are 
usually explained by means of two competing scenarios, as cited in the Section 1: 1) the triaxiality of the halo 
(\citealt{Newberg06, Xu06, J08}), and 2) the remnants of some historical merger events (\citealt{Wyse05}). Fig. 18 shows 
that the maximum and minimum surface densities for the apparent magnitudes $18<g_{0}\leq22$, corresponding to halo component 
of the Galaxy, do not fit with the galactic longitudes $l=0^{\circ}$ and $l=180^{\circ}$, respectively, confirming these scenarios. 
Instead, the maximum surface density is at $l=20^{\circ}$. But, this value is close to the direction, which the long disc bar points out. 
This finding encouraged us to argue that the slight minimum and the small maximum cited above originate from the long bar 
effect. To confirm this argument, one needs to remember that the maximum of $(c/a)$ corresponds to the fields in the third quadrant, 
i.e. $210^{\circ}\leq l \leq 220^{\circ}$ where long bar lies, whereas its minimum is obtained for the star fields 
deficient in the second quadrant. Also, we should remind that we worked on intermediate latitude fields, 
where the halo is rather shallow. The minimum and maximum accretions, i.e. Sagittarius streams (\citet{Newberg02}, their Fig. 1) 
and/or Palomar 5 \citep{Oden03} are centred at the galactic coordinates ($0^{\circ},45^{\circ}$) and ($20^{\circ},45^{\circ}$), 
respectively. 

Three excesses of surface densities, ($l=20^{\circ}$), ($l=50^{\circ}$, $19<g_{0}\leq20$), ($l=210^{\circ}$, $20<g_{0}\leq22$), 
in Fig. 18 correspond to the overdensity regions cited in the literature i.e. Herculus-Aquila Cloud \citep{Belokurov07}, and 
Sagittarius Star Stream \citep{Newberg06}. Thus, the distribution of surface densities in 
Fig. 18 summarizes many properties of the galactic components. 

\subsection{Metallicity Gradient in the Vertical Direction}

The metallicity distribution could be obtained up to distances $z=10$ kpc from the galactic plane, which covers the thin 
and thick discs and inner halo. There is an agreement between the metallicity gradients in the vertical direction evaluated 
in our work and the corresponding ones appearing in the literature. That is, for short distances, the metallicity gradient 
is $d[M/H]/dz \sim-0.3$ dex kpc$^{-1}$, confirming the formation of the thin disc by dissipational collapse. The metallicity 
gradient is steeper in the transition regions of two galactic components, such as thin and thick discs, and finally the inner 
halo may be assumed as a component of zero metallicity gradient, which is subject to a different formation scenario, i.e. one 
expects a contamination from mergers or accretion of numerous fragments from objects such as dwarf galaxies.        

\section{Acknowledgments}
We thank the anonymous referee for a thorough report and useful comments that helped improving an early version of the paper.
We would like to thank Dr. Sel\c cuk Bilir, Dr. Serap Ak, Dr. Antonio Cabrera-Lavers and K. Ba\c sar Co\c skuno\u glu for their 
contributions. Also, we thank to Hikmet \c{C}akmak and Tu\u{g}kent Akkurum for preparing some computer codes for this study. 
Salih Kaarali thanks to Beykent University for financial support.

The {\em SDSS} is managed by the Astrophysical Research Consortium (ARC) for the Participating Institutions. 
The Participating Institutions are The University of Chicago, Fermilab, the Institute for Advanced Study, 
the Japan Participation Group, The Johns Hopkins University, the Korean Scientist Group, Los Alamos National 
Laboratory, the Max-Planck-Institute for Astronomy (MPIA), the Max-Planck-Institute for Astrophysics (MPA), 
New Mexico State University, University of Pittsburgh, University of Portsmouth, Princeton University, the 
United States Naval Observatory, and the University of Washington. Funding for the creation and distribution 
of the {\em SDSS} Archive has been provided by the Alfred P. Sloan Foundation, the Participating Institutions, 
the National Aeronautics and Space Administration, the National Science Foundation, the U.S. Department of 
Energy, the Japanese Monbukagakusho, and the Max Planck Society. The {\em SDSS} Web site is http://www.sdss.org/.

\end{document}